\documentclass[doublecol]{epl2}
\pdfoutput=1
\usepackage[T1]{fontenc}
\usepackage[latin9]{inputenc}
\usepackage{textcomp}
\usepackage{amsmath}
\usepackage{amssymb}
\usepackage{graphicx}
\usepackage{esint}
\usepackage{color,soul}
\usepackage[english]{babel} 
\usepackage{xcolor}
\usepackage{soul}

\title{The magnetic Purcell effect: the case of an emitter near an antiferromagnet }
\author{ Beatriz A. Ferreira\inst{1,2} \and N. M. R. Peres\inst{1,2}}
\institute{
	\inst{1} Departamento e Centro de F\'isica, Universidade do Minho - 4710-057, Braga, Portugal\\
	\inst{2} International Iberian Nanotechnology Laboratory (INL) - 4715-330, Braga, Portugal
	
	
}

\abstract{%
  In this paper we discuss the magnetic Purcell effect of a magnetic dipole near a semi-infinite antiferromagnet. Contrary to the electric Purcell effect, the magnetic one  is not so well studied in the literature. We derive the dispersion relation of the 
  surface wave existing at an  antiferromagnetic-dielectric interface from the calculation of the reflection coefficient of the structure. After characterizing the surface wave we quantize the 
  electromagnetic vector potential of the surface wave.
  This allow us to discuss the magnetic Purcell effect via Fermi golden rule.
}
\begin{document}
\maketitle
\section{Introduction}
\label{intro}

The electric Purcell effect \cite{purcell1946} refers to the modification of the lifetime of an electric dipolar transition, relative to its value in vacuum, when an emitter (including, for example, atomic and molecular transitions, and quantum dots) is positioned near dielectric or metallic bodies, including periodic structures \cite{purcell}, microcavities \cite{vahala2003}, nanoantennas \cite{novotny2011,hein2013},
metamaterials \cite{yuri2013},  nanoparticles \cite{chigrin2016,wiecha2018,liang2019}
and 2D materials \cite{book,kamp2015}.
The magnetic Purcell effect is the counterpart of that when a magnetic dipolar transition is involved.

In general terms, when an emitter has both 
electric and magnetic dipolar transitions, the emitter decays preferentially via the electric dipolar transition, thus obscuring the magnetic one. If we take the ratio of the magnetic energy interaction of a magnetic dipole to the electric energy interaction of an electric dipole we obtain a number proportional to the fine structure constant, $\alpha\approx 1/137$. This explains the weakness of the magnetic dipolar transtion. In addition, the interaction of magnetic dipoles with the environment is weak \cite{wang2017}, because relative magnetic permeability of common materials, such as dielectrics, is small, approximately 1.  There are however cases where the interaction can be enhanced \cite{kivshar2015}.
 There are however cases (for example, rare-earth ions \cite{hussain2015} and semiconductor quantum dots)  that either the electric dipolar transition is forbidden or both dipolar electric and magnetic transitions are equally preferable \cite{zurita2002,baranov2017,li2018,hernandes2018,paz2018}. Because
dipolar magnetic transitions are, in general, not the dominant
 electromagnetic transitions,
 the literature on the magnetic Purcell effect is scarce \cite{soljacic2018,feng2018,slobozhanyuk2014,wu2019}.
Another case where the magnetic dipole transition can be enhanced happens when a the magnetic dipole is located in the vicinity of a body with a large magnetic permeability. Such condition can occur naturally when the body is a magnetic material (ferromagnetic or antiferromagnetic materials near the spin wave resonance)  or when the body is a metamaterial specifically designed for having a strong magnetic response. The first case can take place at frequencies 
from the gigahertz to the terahertz whereas the second case can happen in frequencies as low as the 
microwaves (gigahertz).

In this paper we study the magnetic Purcell effect, that is, the modification of the decay rate of a magnetic dipolar transition due to the presence of a magnetic body (an antiferromagnet). We show below that a surface wave
\cite{cottam2015}
 exists is a narrow region in the momentum-frequency space. In that region the decay rate of the magnetic dipolar transition varies by orders of magnitude, specially when the surface wave becomes strongly localized in space.
In order to compute the modification of the decay rate 
we quantize the electromagnetic field and use Fermi golden rule.
\section{Dispersion relation of the surface wave near an antiferromagnet-dielectric interface}
\label{sec-1}

\begin{figure}
\centering
\includegraphics[width=8cm,clip]{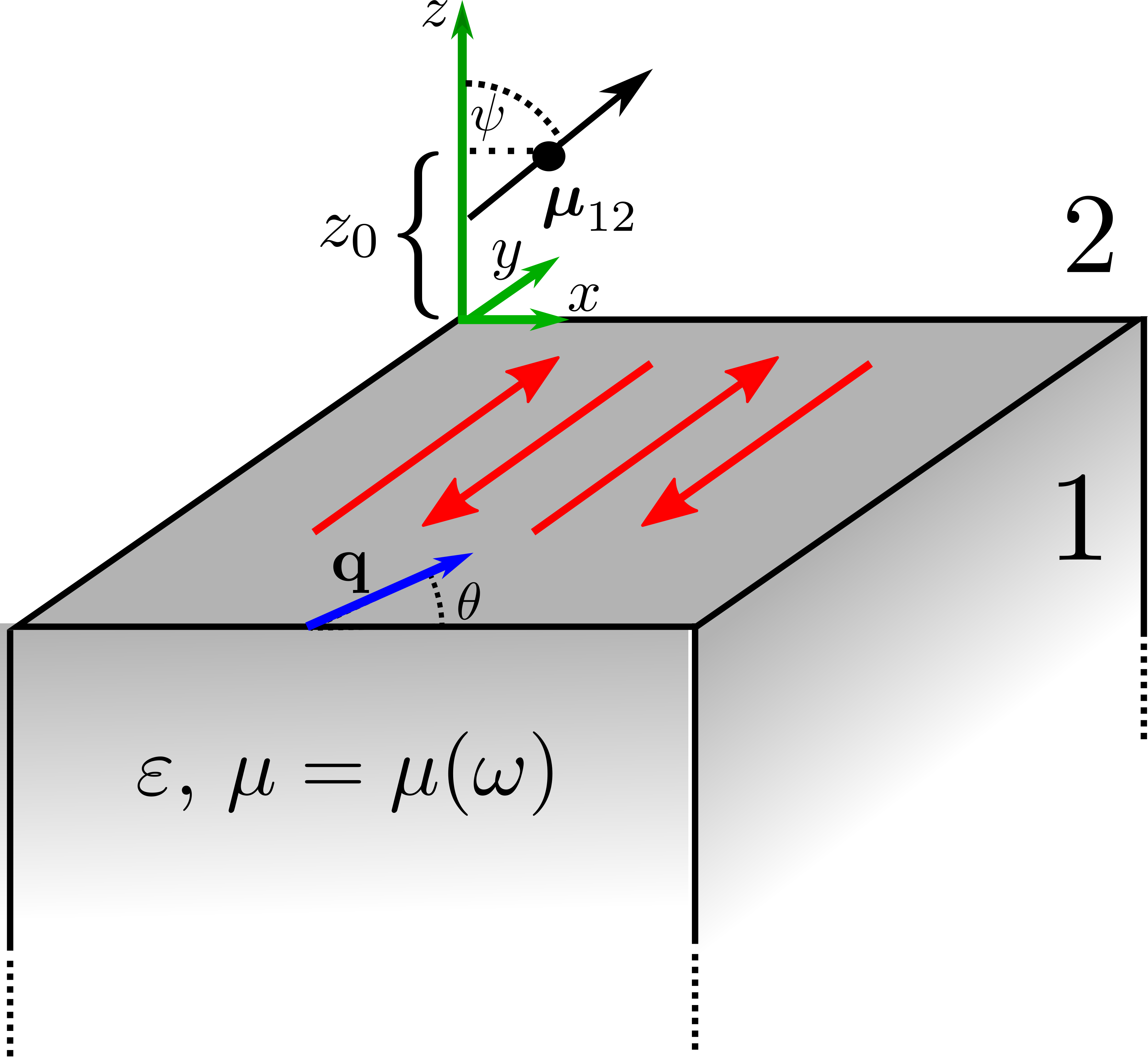}
\caption{Geometry of the system. An antiferromagnet with a staggered magnetization along the $y-$direction occupies the half-space $z<0$ (medium $j=1$). A magnetic dipole, $\boldsymbol{\mu}_{12}$, lies in the half-space $z>0$ at a distance $z_0$ from the surface of the antiferromagnet oriented with an angle $\psi$ in relation to the $z$-axis. The staggered magnetization lies in the $xy-$plane. The momentum vector, $\mathbf{q}=(q_x,q_y)$ makes an angle $\theta$ with the $x$-axis. The decay of the excited magnetic dipole induces surface waves propagating at the interface of the antiferromagnet.}
\label{fig-1}       
\end{figure}

In this section we derive the dispersion relation of a surface wave existing at a dielectric-magnetic interface. To be definitive, we use  an antiferromagnet as the magnetic body (see Fig. \ref{fig-1}) whose staggered magnetization lies in the $xy-$plane. 
In Fig. \ref{fig-1} the magnetic dipole will decay and excited an eletromagnetic wave surface wave --a polariton-- coupled to the magnetic excitations of the antiferromagnet.
The goal is to compute the reflection coefficient of this structure, from where the dispersion relation of the surface wave existing at the dielectric-antiferromagnet system can be derived. 
For determining the fields, we need to solve Maxwell equations:
\begin{equation}
\boldsymbol{\nabla}\times \mathbf{E}=-\frac{\partial \mathbf{B}}{\partial t},
\label{eq:Max11}
\end{equation}
\begin{equation}
\boldsymbol{\nabla}\times\mathbf{H}=\frac{\partial \mathbf{D}}{\partial t},
\label{eq:Max22}
\end{equation}
We assume that the eletric field can be written in medium 1 as 
\begin{align}
\mathbf{E}_1=(E_{1,x},E_{1,y},0)e^{i(\mathbf{q}\cdot\boldsymbol{\rho}-\omega t)} e^{-i\beta_1z}\nonumber\\
+(-E_{1,x},E_{1,y},0)r_se^{i(\mathbf{q}\cdot\boldsymbol{\rho}-\omega t)} e^{i\beta_1z}
\label{3}
\end{align}
and in medium 2 as
\begin{equation}
\mathbf{E}_2=(E_{1,x},E_{1,y},0)t_s e^{i(\mathbf{q}\cdot\boldsymbol{\rho}-\omega t)} e^{-i\beta_2z}
\label{4}
\end{equation}
with $\beta_j=\sqrt{\omega^2\varepsilon_j\mu_j/c^2-q^2}$, where $c^2=(\mu_0\varepsilon_0)^{-1}$ is the speed of light in vacuum, $\mathbf{q}$ and $\bm{\rho}$ are 2D vectors in the $xy-$plane representing the 2D wave vector and the 2D position vector, respectively, $r_s$ and $t_s$ are the Fresnel reflection and transmission coefficients for the TE polarization, respectively. The magnetic field can be obtain from equation(\ref{eq:Max11})  \text{using equations} (\ref{3})  \text{and} (\ref{4}).
For simplicity we choose the polarization of the eletric field along the $y$-axis.Therefore, the wave equation for this component of the electric field reads 
\begin{equation}
\nabla ^2 E_y(x,z)-\mu_j\mu_0 \varepsilon_j\varepsilon_0 \frac{\partial ^2 E_y(x,z)}{\partial t^2}=0,
\label{Rot_Far}
\end{equation}
where  $\varepsilon_j$ is  the relative dielectric permittivity  of the medium $j$, $\varepsilon_0$ is the dielectric permittivity of vacuum, $\mu_j$ is the relative permeability of the medium $j$, and  $\mu_0$ is the magnetic permeability of the vacuum.We assume a harmonic time dependence for the eletromagnetic field in the form $e^{-i\omega t}$ and we obtain
\begin{equation}
\left(\frac{d^2}{dz^2}-q^2 +\omega^2\varepsilon_j\mu_j\varepsilon_0\mu_0\right) E_y(q,z)=0.
\label{eq_dif}
\end{equation}
Using the constitutive relation
 $\mu_0\mu_j\mathbf{H}=\mathbf{B}$, we can write the magnetic field in medium 2 as:
\begin{align}
&\mu_0\mathbf{H}^{(2)}(q,z)=\hat{z}\frac{i}{\omega\mu_2}iq[E_{1,y}e^{-i \beta_2 z}+ E_{1,y}r_se^{i\beta_2 z}] e^{i(q x-\omega t)}\nonumber\\
&-\hat{x}\frac{i}{\omega\mu_2}[-i\beta_2 E_{1,y}e^{-i \beta_2 z}+ i\beta_2E_{1,y}r_se^{i\beta_2 z} ] e^{i(q x-\omega t)}
\label{5}
\end{align}
and in medium 1 as:
\begin{align}
\mu_0\mathbf{H}^{(1)}(q,z)&=\frac{i}{\mu_1(\omega)\omega}[i\beta_1E_{1,y}t_se^{-i\beta_1 z}\hat{x}\nonumber\\
&+iqE_{1,y}t_se^{-i\beta_1 z}\hat{z}]e^{i(q x-\omega t)}.
\label{6}
\end{align}
Next we connect the fields in the two regions using the boundary conditions

\begin{align}
\mathbf{E}^{(1)}_t(q,0)&=\mathbf{E}^{(2)}_t(q,0)\\
\mathbf{H}^{(1)}_t(q,0)&=\mathbf{H}^{(2)}_t(q,0),
\label{c1}
\end{align}
where the index $t$ refers to the tangential component of the fields.
\text{Substituting Eqs.} (\ref{3})-(\ref{4}) \text{and} (\ref{5})-(\ref{6}) in the boundary conditions we obtain
\begin{align}
1+r_s&=t_s\\
\frac{1}{\mu_1(\omega)}\beta_1 t_s&=\frac{1}{\mu_2}(\beta_2-\beta_2r_s).
\end{align}
The previous linear system can be easily solved and we obtain
\begin{equation}
r_s=-\frac{\mu_2\beta_1-\mu_1(\omega)\beta_2}{\mu_2\beta_1+\mu_1(\omega)\beta_2},,
\label{loss}
\end{equation}
As usual, the equation giving the dispersion relation of the surface wave follows from the poles of $r_s$, that is, from the condition
\begin{equation}
\mu_2\beta_1+\mu_1(\omega)\beta_2=0,
\label{eq_dispersion}
\end{equation}
a result previously derived in the literature using a different method \cite{camley1982}. The solutions of the previous equation exist in a narrow energy range (see below), near the spin wave resonance where the relative magnetic permeability is negative.

\begin{figure}
\centering
\includegraphics[width=8.5cm,clip]{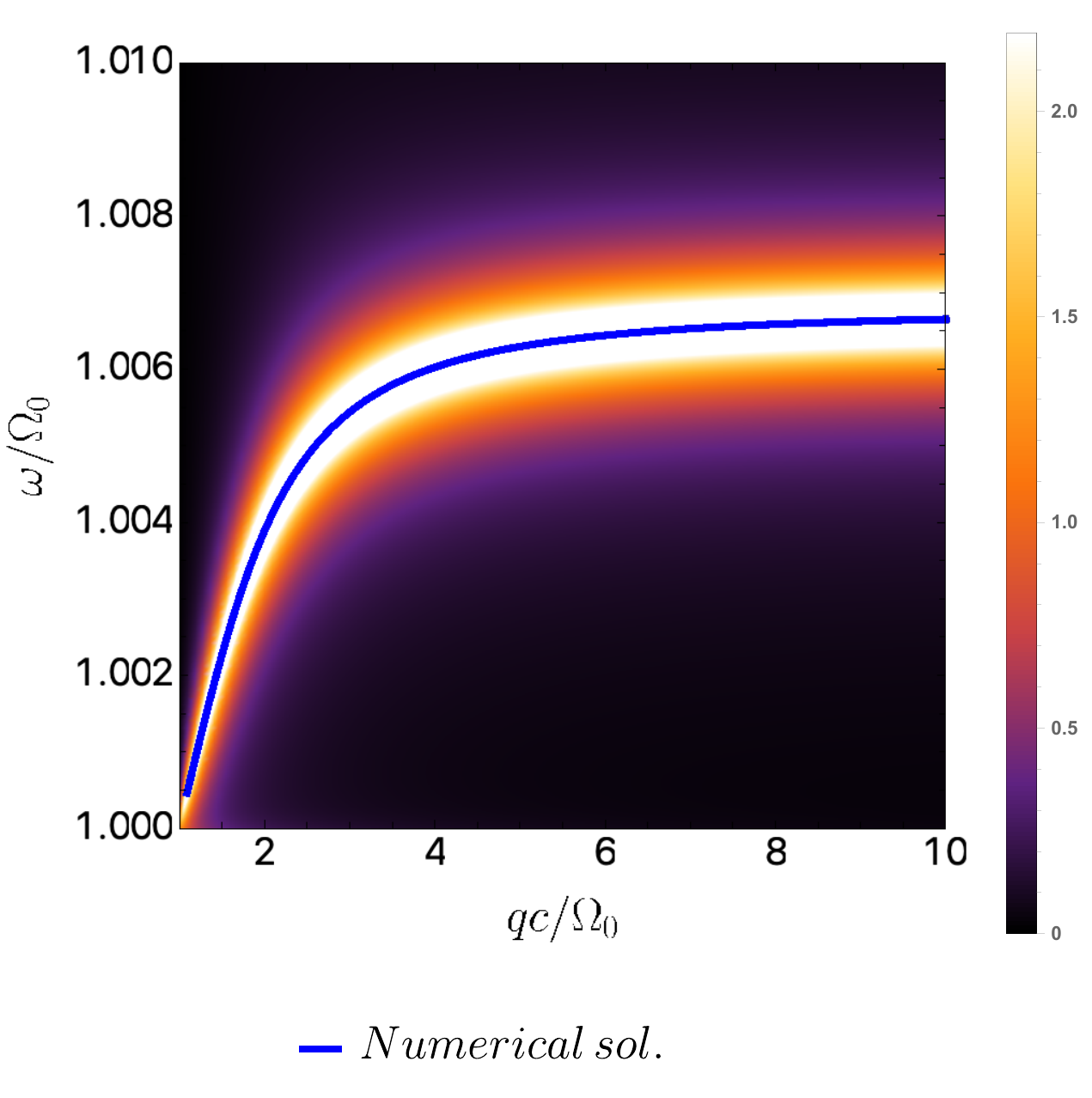}
\caption{Dispersion relation of the surface wave. 
The density plot represents the loss function. The blue  line represents the solution of Eq. 
(\ref{eq_dispersion}), that is, the dispersion relation of the surface wave. The relative permittivity of the dielectric constant of the antiferromagnet was chosen as $\epsilon_1=6$.}
\label{fig-2}       
\end{figure}

The loss function is defined as minus the imaginary part of the reflection amplitude:
\begin{equation}
\mathcal{L}=-\Im(r_s).
\end{equation}
For an antiferromagnet $\mu_1(\omega)$ is given by \cite{kittel1952}
\begin{equation}
\mu_1(\omega)=1+\frac{2 \Omega_s^2 }{\Omega_0^2-\left( \omega +i\Gamma_r\right)^2},
\label{eq_mu}
\end{equation}
where $\Gamma_r=1/\tau$ is the relaxation rate, $\Omega_0=\gamma \mu_0 \sqrt{H_a^2+2 H_a H_e}$ is the antiferromagnetic resonance frequency, $\Omega_s=\gamma \mu_0\sqrt{2H_aM_s}$ is the saturation frequency,  $\gamma=e/(2m)$ is the giromagnetic ratio, $e$ is the elementary charge, and $m$ is the electron mass. The quantities $H_a$, $H_e$, and $M_s$ are given is Table \ref{tab-1}. The solutions of Eq. (\ref{eq_dispersion}) exist in the range
$\Omega_0<\omega<\sqrt{\Omega_0^2+2\Omega_s^2}$, which, for the parameters of Table \ref{tab-1}, fall in the THz spectral range. The spectrum of the surface wave is given in Fig. \ref{fig-2} (note the magnitude of the vertical scale). For energies close to $\Omega_0$
the dispersion merges with the light line.  Therefore, the surface wave is poorly localized in space. For large wave numbers (small wave lengths), the dispersion is almost flat and the surface wave is strongly localized in space.

\begin{table}
\caption{Parameters characterizing the
MnF$_2$
 antiferromagnet \cite{macedo}.
The frequency $\Omega_0$ is the frequency 
 of the antiferromagnetic resonance and $\tau=1/\Gamma_r$ is the relaxation time.}
\label{tab-1}       
\begin{tabular}{|l|l|l|l|l|}
\hline
$\mu_0H_a$ & $\mu_0H_e$ & $\mu_0M_s$    & $\Omega_0$ ($\times 10^{12}$)& $\tau$ ($\times 10^{-9}$)\\\hline
0.787 T & 55.3 T & 0.756 T &   1.69 rad/s& 7.58 s \\
\hline
\end{tabular}
\vspace*{0cm}  
\end{table}

\section{Quantization of the electromagnetic field}
\label{sec-2}
In this section we show how to quantize the electromagnetic field of the surface wave and derive the quantum mechanical version of the electromagnetic energy. The section is divided in two parts: the calculation of the electromagnetic fields of the surface wave whose spectrum was determined in the previous section, and the quantization of the vector potentential of the electromagnetic field (we work in Weyl gauge, where the electrostatic potential is zero).  
Our eletric and magnetic field will be similar to the one presented in equations (\ref{3}) and (\ref{4}) 

\begin{align}
\mathbf{E}_j&=(E_{j,x},E_{j,y},0)e^{i(\mathbf{q}\cdot\boldsymbol{\rho}-\omega t)} e^{-\kappa_j|z|}
\label{20}\\
\mathbf{H}_j&=\frac{1}{\mu_j}(H_{j,x},H_{j,y},H_{j,z})e^{i(\mathbf{q}\cdot\boldsymbol{\rho}-\omega t)} e^{-\kappa_j|z|},
\label{21}
\end{align}	
but here we removed the incident field and consider $\beta_j=i\kappa_j$. This will turn our waves into evanescent waves with
\begin{equation}
\kappa_j^2=q^2-\frac{\varepsilon_j\mu_j \omega^2}{c^2}.
\end{equation}
and the dispersion of the surface wave given by
the expression found from the pole of the reflection coefficient. The relation between the amplitudes of the fields read
\begin{align}
B_{j,x}&=\frac{i(-1)^j\kappa_j}{\omega}E_{j,y},\\
B_{j,y}&=-\frac{i(-1)^j\kappa_j}{\omega}E_{j,x},\\
B_{j,z}&=\frac{1}{\omega}[q_xE_{j,y}-q_yE_{j,x},]\\
q_xE_{j,x}&=-q_yE_{j,y}.
\end{align}
From these relations the electric field takes the form ($q_x\ne 0$ and $q_y\ne 0$)
\begin{equation}
\mathbf{E}_j(\mathbf{r},t)=E_{j,x}\left(\hat{x}-\frac{q_x}{q_y}\hat{y}\right)e^{i(\mathbf{q}\cdot\boldsymbol{\rho}-\omega t)}e^{-\kappa_j |z|}.
\label{28}
\end{equation}
The electric field can be obtained from the vector potential as
\begin{equation}
\mathbf{A}(\mathbf{r},t)=-\frac{\partial \mathbf{E}(\mathbf{r},t)}{\partial t}\\.
\end{equation}
We can write the real vector potential as superposition of electromagnetic modes:
\begin{align}
\mathbf{A}_j(\mathbf{r},t)&=\sum_{\mathbf{q}}[A_\mathbf{q}\mathbf{u}^j_\mathbf{q}(z)e^{i\mathbf{q}\cdot\boldsymbol{\rho}}e^{-i\omega_{sm}t}
\nonumber\\
&+A^{*}_\mathbf{q}[\mathbf{u}^{j}_\mathbf{q}(z)]^\ast e^{-i\mathbf{q}\cdot\boldsymbol{\rho}}e^{i\omega_{sm}t}],
\label{eq:vec_potential}
\end{align}
where $\mathbf{u}^j_\mathbf{q}(z)$ is called the mode function and $\omega_{sm}=\omega_{sm}(\mathbf{q})$ is the frequency of the surface wave. From the form of the electric field,
we write the mode as
\begin{equation}
\mathbf{u}^j_\mathbf{q}(z)=\frac{1}{\sqrt{L}}\left(\hat{x}-\frac{q_x}{q_y}\hat{y}\right)e^{-\kappa_j |z|},
\label{eq_mode}
\end{equation}
where $L$ is a constant to be determined latter and is called the mode length.

The quantization procedure starts with the classical form of the electromagnetic energy contained in the field of the surface wave. The energy is given by the usual expression
(assuming a constant dielectric function)
\cite{ruppin2002,tame2013,milonni2019}

\begin{align}
U_f&=\int d^3\mathbf{r}\left[\frac{\varepsilon_0\varepsilon_j}{2} \mathbf{E}^2(\mathbf{r},t)\right.
\nonumber\\
&+\left.\frac{\mu_0}{2 }\mathbf{H}(\mathbf{r},t)\frac{d}{d \omega}\left(\omega\bar{\bar\mu}_j(\omega)\right)\mathbf{H}(\mathbf{r},t)\right]\,,
\label{eq_U}
\end{align}
 where $\bar{\bar\mu}_j(\omega)$ is the relative magnetic permittivity tensor, given by
 \begin{equation}
\bar{\bar\mu}_1(\omega)=\begin{bmatrix}
\mu_1(\omega)& 0&0\\
0&1&0\\
0&0&\mu_1(\omega)
\end{bmatrix}\,,
\label{eq_mu_tensor}
\end{equation}
and where $\mu_1(\omega)$ is given by 
Eq. (\ref{eq_mu}) and $\bar{\bar\mu}_2(\omega)$ is a unit matrix of dimension 3.

After lengthy calculations and demanding that the energy in the field has the form

\begin{equation}
U_f=S \varepsilon_0\sum_{\mathbf{q}} \omega_{sm}^2[A_\mathbf{q}A^\ast_\mathbf{q}+A^\ast_\mathbf{q}A_\mathbf{q}],
\label{eq_Uf_harmonic}
\end{equation}
where $S$ is the area of the antiferromagnetic surface in the $xy-$plane, it follows the mode length $L$ as

\begin{equation}
L=\sum_{j=1}^2\frac{1}{2\kappa_j q^2_y}\left(\varepsilon_jq^2+\frac{\bar\mu_jq_x^2\kappa_j^2+q_y^2\kappa_j^2+\bar\mu_jq^4}{2\varepsilon_0\mu_0\omega_{sm}^2}\right)\,,
\label{eq_L}
\end{equation}
where $q_x=q\cos\theta$, $q_y=q\sin\theta$, 
\begin{equation}
\bar\mu_1=\frac{d (\omega\mu_1)}{d\omega}\,,
\end{equation}
and $\bar\mu_2=1$.

Next we quantize the Hamiltonian making the transformations
\begin{equation}
A_\mathbf{q}\rightarrow\sqrt{\frac{\hbar}{2S\varepsilon_0\omega_{sm}}}a_\mathbf{q}\,,
\label{eq_A}
\end{equation}
and
\begin{equation}
A^*_\mathbf{q}\rightarrow\sqrt{\frac{\hbar}{2S\varepsilon_0\omega_{sm}}}a^\dagger_\mathbf{q},
\label{Eq_Adagger}
\end{equation}
where $a_\mathbf{q}$ and $a^\dagger_\mathbf{q}$ are second quantized operators obeying the usual canonical commutation relation
 $[\mathbf{a}_{\mathbf{q}_1},\mathbf{a}^\dagger_{\mathbf{q}_2}] =\delta_{\mathbf{q}_1,\mathbf{q}_2}$. These substitutions lead to the second quantized harmonic oscillator Hamiltonian
 \begin{equation}
H=\frac{1}{2}\sum_{\mathbf{q}}\hbar\omega_{sm}[a_\mathbf{q}a^\dagger_\mathbf{q}+a^\dagger_\mathbf{q}a_\mathbf{q}].
\label{eq_H_quantized}
\end{equation}
In possession of the quantized vector potential we can compute the 
change of the decay rate of an emitter characterized by a magnetic dipolar transition in the presence of a magnetic body, that is, the magnetic Purcell effect.

\section{The magnetic Purcell effect}
\label{sec-3}
In this section we derive the transition rate of an emitter characterized by a magnetic dipolar transition in the vicinity of a magnetic body. For achieving this goal, we use the quantized version of the electromagnetic vector potential (obtained in the previous session) together with Fermi golden rule. For a magnetic dipolar transition characterized by a dipolar magnetic moment $\bm{\mu}_{12}$ the decay rate reads
\begin{equation}
\Gamma=\frac{2\pi}{\hbar}\sum_{\mathbf{q}}\vert \langle 1;n_\mathbf{q}+1\vert\bm{\mu}_{12}\cdot\mathbf{B}\vert 2;n_\mathbf{q}\rangle\vert^2
\delta(\hbar\omega_{at}-\hbar\omega_{sm})
\label{eq_Fermi}
\end{equation}
where $\hbar\omega_{at}$ is the energy of the atomic transition
and  $\vert j;n_\mathbf{q}\rangle$ is the state of the system where the 
atom is in the state $j$ and with  $n_\mathbf{q}$ surface waves
present. We shall consider the simplest case of a transition of the form
$\vert 2;0\rangle\rightarrow\vert 1;1\rangle$. This corresponds to the emitter being initially  in the excited state and no surface wave is present followed by a transition to the ground state of the emitter with the corresponding excitation of a surface wave of wavevector $\mathbf{q}$.
We also note that the transition rate in vacuum is given by \cite{baranov2017}
\begin{equation}
\Gamma_0=\frac{\mu_0\omega^3_{at}}{3\pi \hbar c^3}\bm{\mu}^2_{12},
\end{equation}
where $\bm{\mu}_{12}=\mu_{12}(\sin\psi,0,\cos\psi)$ and $\psi$ is the angle the magnetic dipole makes with the $z-$axis.
\begin{figure}
\centering
\includegraphics[width=8.5cm,clip]{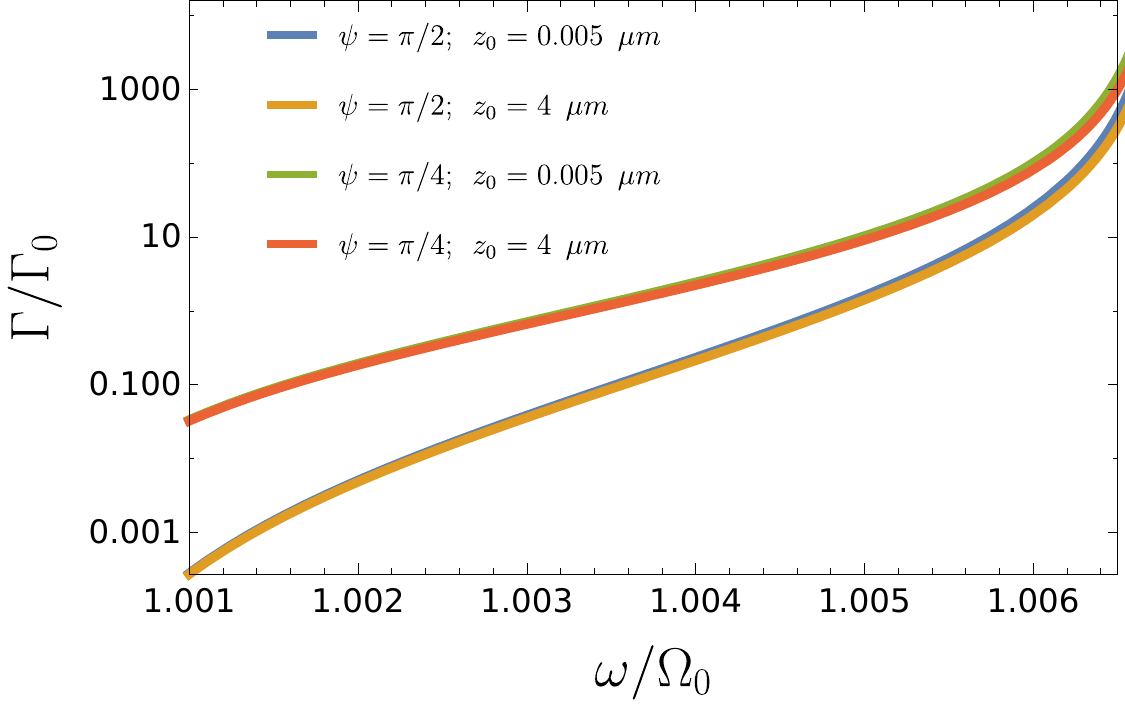}
\caption{Magnetic Purcell factor for  an emitter at different distances from the surface of an antiferromagnet and at different angles from the perpendicular to the antiferromagnetic interface. The Purcell factor diverges as we approach the frequency
$\sqrt{\Omega_0^2+2\Omega_s^2}$, a consequence of ignoring losses in the antiferromagnet.}
\label{fig-3}       
\end{figure}
For computing $\Gamma$ we need the matrix element of the magnetic energy, with the magnetic field written in second quantization. 
The matrix element reads:

\begin{equation}
\langle 1;1\vert\bm{\mu}_{12}\cdot\mathbf{B}\vert 2;0\rangle\
=\sqrt{\frac{\hbar}{2S \epsilon_0 \omega_{sm}}}\bm{\mu}_{12}\cdot[\nabla	\times\mathbf{u}^*_\mathbf{q}(z)e^{-i\mathbf{q}\cdot\bm{\rho}}]\,.
\end{equation}
Once the curl in the matrix element is computed, the transition rate follows as (the integration of the $\delta-$function in Eq. (\ref{eq_Fermi}) is elementary):
\begin{align}
\Gamma	&=\frac{\mu_{12}^{2}\hbar}{4\pi\epsilon_{0}}\int_{0}^{2\pi}d\theta q(\hbar\omega_{at})B(\hbar\omega_{at})\frac{e^{-2\kappa_{2} z_0}}{\hbar\omega_{at}L(\omega_{at},\theta)\sin^2\theta}
\nonumber\\
&\times
\left(\kappa_2^2\cos^2 \theta \sin^2 \psi+q^2 \cos^2 \psi\right)\,,
\end{align}
where $q(\hbar\omega_{at})$ follows from the dispersion of the surface wave computed in
Eq. (\ref{eq_dispersion}) and reads
\begin{equation}
q(\hbar\omega)=\frac{\hbar\omega\sqrt{(\hbar\Omega_0)^2+2(\hbar\Omega_s)^2-(\hbar\omega)^2}}{\hbar c\sqrt{2}\sqrt{(\hbar\Omega_0)^2+(\hbar\Omega_s)^2-(\hbar\omega)^2}}\,,
\label{eq_q_w}
\end{equation}
and $B(\hbar\omega)$ is defined as 
$dq=B(\hbar\omega)d(\hbar\omega)$.
We have also made explicit the dependence
of $L=L(\omega_{at},\theta)$ on $\theta$ and 
$\omega_{at}$.
\begin{figure}
\centering
\includegraphics[width=8.5cm,clip]{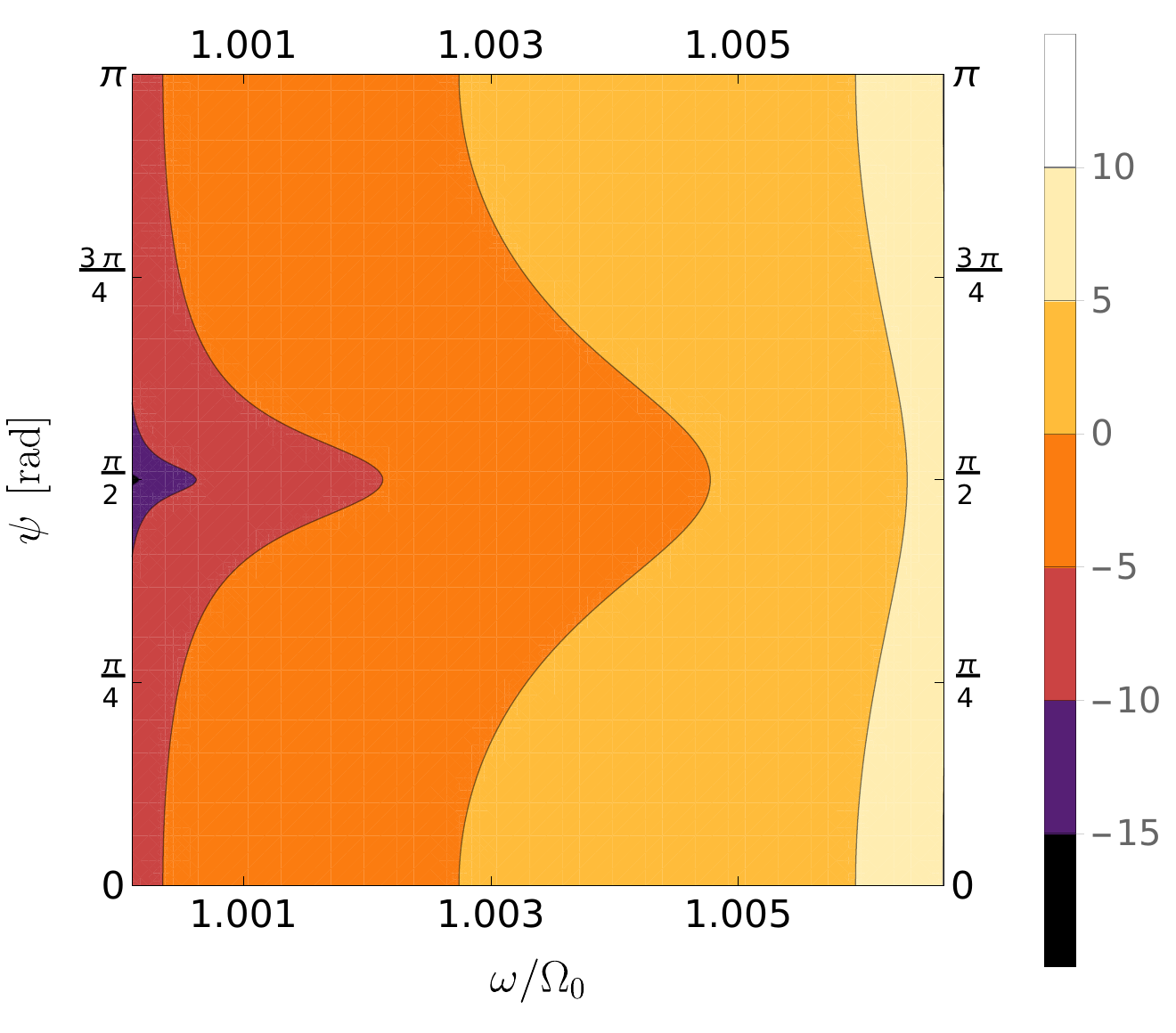}
\caption{Contour density plot of the magnetic Purcell factor as function of $\omega/\Omega_0$
and $\psi$, for $z_0=4$ nm. For clarity, we have represented $\ln(\Gamma/\Gamma_0)$.}
\label{fig-4}       
\end{figure}

In Fig. \ref{fig-3} we represent the magnetic Purcell factor for two different distances $z_0$ of the dipole to the antiferromagnetic surface
and two different orientations $\psi$ of the magnetic dipole relatively to the $z-$axis. 
For a dipole parallel to the antiferromagnetic surface ($\psi=\pi/2$) the Purcell factor is smaller than when the dipole is at an angle ($\psi=\pi/4$ in this case). Also, we see that the Purcell factor can vary over 6 orders of magnitude, staring at values smaller than 1 up to values of the order of 1000.
The smallest Purcell factor occurs when the frequency of emission is close to $\Omega_0$ and it increases from there onward. The increase of the Purcell factor is linked to the degree of  localization of the surface wave. The more the surface wave is localized (larger values of $q$; see Fig. \ref{fig-2}) the larger is the Purcell factor.
If  disorder is taken into account there will be a $q^\ast$ where the dispersion seen in Fig. \ref{fig-2} folds back. This point defines a frequency $\omega^\ast$. Above the energy $\hbar\omega^\ast$ the Purcell factor decreases because the surface wave becomes over damped.

In Fig. \ref{fig-4} we provide a contour density plot
of the Purcell factor as function of energy and the angle $\psi$. Clearly for $\psi=\pi/2$ the magnetic Purcell factor has its lowest value and is symmetric relatively to that point, a consequence of the dependence of $\Gamma$ on the square of the trigonometric functions of $\psi$.
\section{Conclusions}
\label{sec-conclu}

In this paper we have analyzed the magnetic Purcell effect. We have considered an emitter, characterized by a magnetic dipolar transition, in the vicinity of a magnetic body. We found that the decay of the emitter 
is enhanced by orders of magnitude when the frequencies of the surface wave correspond to highly localized states (large wave numbers $q$). When the 
frequency tends to $\Omega_0$ the dispersion merges with the light line and the surface wave becomes poorly localized in space. In this case the transition rate is suppressed with $\Gamma/\Gamma_0<1$.
An extension of this work is to consider a system where a graphene sheet is kept at a fixed distance from the antiferromagnet surface. 
In this geometry it as been shown by one of us (NMRP) that doping graphene induces a substantial change in the dispersion of the surface wave. This is an additional control on the spectral position of the dispersion of the surface wave and thus also over the  ratio
$\Gamma/\Gamma_0$.

\section{Appendix}

In this appendix we give an example of how to express the energy in field
in terms of amplitudes $A_{\mathbf q}$ and $A_{\mathbf q}^\ast$. To that end let us consider the contribution coming from the electric field:

\begin{align}
&\int d^3\mathbf{r}\frac{\varepsilon_0\varepsilon_j}{2}  \mathbf{E}^2(\mathbf{r},t)=\int dz \int d^2\boldsymbol{\rho}\frac{\varepsilon_0\varepsilon_j}{2} \mathbf{E}^2(\mathbf{r},t)=\nonumber\\
&\frac{\varepsilon_0\varepsilon_j}{2}\int dz S \sum_{\mathbf{q}}\omega_{sm}^2[A_\mathbf{q}A^\ast_\mathbf{q}+A^\ast_\mathbf{q}A_\mathbf{q}] u^j_\mathbf{q}(z)\cdot [u_\mathbf{q}^{j}(z)]^\ast=\nonumber\\
&\frac{\varepsilon_0}{2}S \sum_{\mathbf{q}}\omega_{sm}^2[A_\mathbf{q}A^\ast_\mathbf{q}+A^\ast_\mathbf{q}A_\mathbf{q}]\left(\int_{-\infty}^0 dz \varepsilon_2 u^2_\mathbf{q}(z)\cdot [u_\mathbf{q}^{2}(z)]^\ast\right.
\nonumber\\
&\left.+\int^{\infty}_0 dz \varepsilon_1 u^1_\mathbf{q}(z)\cdot [u_\mathbf{q}^{1}(z)]^\ast\right)=\nonumber\\
&S\sum_{\mathbf{q}}\omega_{sm}^2[A_\mathbf{q}A^\ast_\mathbf{q}+A^\ast_\mathbf{q}A_\mathbf{q}]\sum_j\frac{\varepsilon_0\varepsilon_jq^2}{2\kappa_j Lq_y^2}\,,
\label{39}
\end{align}
where we have dropped terms of the form  $A^\ast_\mathbf{q}A^\ast_\mathbf{q}$ and $A_\mathbf{q}A_\mathbf{q}$, because they average to zero over a period
and $S$ is the surface area of the antiferromagnet. The calculation of the energy contribution coming from the magnetic field is performed along the same lines.

\section{Acknowledgments}
B. A. F. and N. M. R. P. acknowledge discussions on the topic of this paper  
with Bruno Amorim and Gil Farias, and  the Center of Physics of the University of Minho and the University of Porto for funding
 in the framework of
the Strategic Financing UID/FIS/04650/2013.
N. M. R. P. acknowledges support
from the European Commission through the project ``Graphene-Driven
Revolutions in ICT and Beyond'' (Ref. No. 785219), COMPETE2020, PORTUGAL2020,
FEDER and the Portuguese Foundation for Science and Technology (FCT)
through project POCI-01-0145-FEDER-028114.

\end{document}